\documentclass[
  twocolumn,
  prl,
  showpacs,
  amsmath,
  amssymb,
  superscriptaddress,
  floatfix
]{revtex4}

\usepackage{bm}
\usepackage{graphicx}

\begin{document}

\title{Charge transport in graphene with resonant scatterers}

\author{M.~Titov}
\affiliation{
 School of Engineering \& Physical Sciences, Heriot-Watt University,
 Edinburgh EH14 4AS, UK
}
\affiliation{
 DFG Center for Functional Nanostructures,
 Karlsruhe Institute of Technology, 76128 Karlsruhe, Germany
}

\author{P.~M.~Ostrovsky}
\affiliation{
 Institut f\"ur Nanotechnologie, Karlsruhe Institute of Technology,
 76021 Karlsruhe, Germany
}
\affiliation{
 L.~D.~Landau Institute for Theoretical Physics RAS,
 119334 Moscow, Russia
}

\author{I.~V.~Gornyi}
\affiliation{
 Institut f\"ur Nanotechnologie, Karlsruhe Institute of Technology,
 76021 Karlsruhe, Germany
}
\affiliation{
 A.~F.~Ioffe Physico-Technical Institute,
 194021 St.~Petersburg, Russia.
}
\affiliation{
 DFG Center for Functional Nanostructures,
 Karlsruhe Institute of Technology, 76128 Karlsruhe, Germany
}

\author{A.~Schuessler}
\affiliation{
 Institut f\"ur Nanotechnologie, Karlsruhe Institute of Technology,
 76021 Karlsruhe, Germany
}

\author{A.~D.~Mirlin}
\affiliation{
 Institut f\"ur Nanotechnologie, Karlsruhe Institute of Technology,
 76021 Karlsruhe, Germany
}
\affiliation{\mbox{
 Institut f\"ur Theorie der kondensierten Materie,
 Karlsruhe Institute of Technology, 76128 Karlsruhe, Germany
}}
\affiliation{
 Petersburg Nuclear Physics Institute,
 188300 St.~Petersburg, Russia.
}
\affiliation{
 DFG Center for Functional Nanostructures,
 Karlsruhe Institute of Technology, 76128 Karlsruhe, Germany
}

\begin{abstract}
The full counting statistics for the charge transport through an undoped
graphene sheet in the presence of strong potential impurities is studied.
Treating the scattering off the impurity in the \textit{s}-wave approximation,
we calculate the impurity correction to the cumulant generating function. This
correction is universal provided the impurity strength is tuned to a resonant
value. In particular, the conductance of the sample acquires a correction of
$16e^2/(\pi^2 h)$ per resonant impurity.
\end{abstract}


\maketitle

Since the discovery of graphene \cite{Novoselov04} its transport properties have
become a subject of intense studies \cite{Novoselov05, GuineaRMP}. The most
remarkable effects arise when the chemical potential is tuned into a close
vicinity of the Dirac point. In particular, a short and wide sample of
\emph{clean} graphene exhibits a pseudo-diffusive charge transport
\cite{Katsnelson}, with the counting statistics equivalent to that of a
diffusive wire \cite{Tworzydlo06Beenakker08rev, Ludwig}. This equivalence has
been confirmed in recent measurements of conductance and noise in ballistic
graphene flakes \cite{Miao07, Danneau08}. In contrast to conventional metals,
ballistic graphene near the Dirac point conducts better when potential
impurities are added \cite{Titov07, Bardarson07, Schuessler09}. Quantum
interference in disordered graphene is also highly peculiar due to Dirac nature
of carriers. In particular, in the absence of intervalley scattering, the
minimal conductivity $\sim e^2/h$ \cite{Novoselov05} is ``protected'' from
quantum localization \cite{OurPapers}.

Strong impurities creating resonances near the Dirac point (``midgap states'')
are one of the most plausible mechanisms limiting the electron mobility and can
be used exploited for functionalization of graphene. As was shown in Refs.\
\cite{OurPRB, Guinea}, such scatterers provide the concentration dependence of
the conductivity, $\sigma \propto n \ln^2 n$, which is consistent with most of
experimental observations. Possible realizations of such scatterers are
vacancies, adsorbed atoms, molecules, or impurity clusters \cite{Lichtenstein,
Geim}. In particular, important example is hydrogen atoms that can be
controllably added to the graphene sample \cite{graphane}. Resonant scattering
may also be intentionally induced by metallic islands deposited on graphene
surface \cite{Bouchiat}. In this case the strength of impurity can be controlled
by a local external gate.

In this paper we study the effect of strong impurities on electron transport in
graphene. We consider the ballistic transport regime in which the sample size
is smaller than the electron mean free path (low impurity concentration). This
model was experimentally implemented in the samples on the SiO$_2$ substrate
\cite{Morpurgo, Miao07, Danneau08}. The ballistic transport is particularly
relevant for suspended samples where much higher mobilities have been achieved
\cite{Bolotin, Andrei}.

\begin{figure}
 \centerline{\includegraphics[width=0.9\columnwidth]{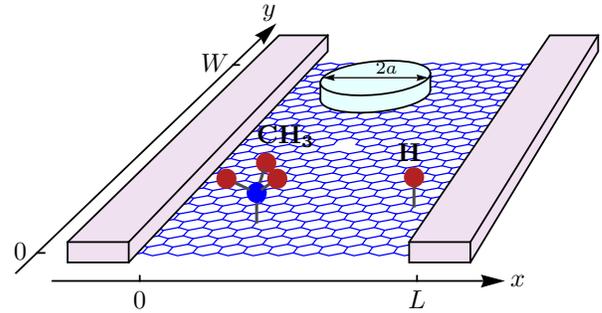}}
 \caption{(Color online) Ballistic graphene setup with various strong
scatterers. Vacancies as well as atomic or molecular impurities can create
midgap states \protect\cite{Lichtenstein}. Metallic islands support quasibound
states that can be tuned to the resonance.}
 \label{Fig:sample}
\end{figure}

We evaluate analytically the effect of strong impurities on the full counting
statistics (FCS) in a rectangular sample of length $L$ and width $W$ (Fig.\
\ref{Fig:sample}). In particular, for the short-and-wide setup, $W \gg L$, we
find the universal corrections to the conductance $G$ and to the shot noise $S$
(divided by $2eV$, where $V$ is the bias voltage),
\begin{equation}
 \label{resonance}
 \delta G
  = \frac{16}{\pi^2} \frac{e^2}{h},
 \qquad
 \frac{\delta S}{2eV}
  = \left(
      \frac{1}{3} - \frac{3}{\pi^2}
    \right) \frac{16}{\pi^2} \frac{e^2}{h},
\end{equation}
per resonant potential scatterer of the round shape.

We consider an effective model described by the Dirac Hamiltonian, $H = -i \hbar
v \bm{\sigma} \mathbf{\nabla} + U(\mathbf{r})$, where $\bm{\sigma} = (\sigma_x,
\sigma_y)$ is the vector of Pauli matrices, $v$ is the velocity, and
$U(\mathbf{r})$ is the impurity potential. Below, we set $\hbar v = 1$. Two
metallic leads at $x < 0$ and $x > L$ are defined by adding a large chemical
potential to the graphene Hamiltonian. Inside the sample, i.e. for $0 < x < L$,
the chemical potential is set to zero (Dirac point). The potential
$U(\mathbf{r})$ represents a collection of randomly distributed sharp scatterers
of arbitrary strength.

We will study FCS of ballistic electron transport through the sample using two
complementary approaches --- scattering theory and Green function formalism
\cite{Nazarov94}. The first approach works for any aspect ratio of the sample
while impurities are treated as point-like. The second method is particularly
suitable for $W \gg L$ and allows for arbitrary impurity profile.

We begin with the scattering approach for a single impurity \cite{Titov07}.
We discretize $y$ coordinate, $y = W \nu / N$, where $\nu = 1, 2, \dots, N$ is
an integer, $N = W/\pi a \gg 1$, and $a \ll L$ plays the role of the impurity
size. The impurity potential in this model is defined as
\begin{equation}
 \label{discrete}
 U(\mathbf{r})
  = \alpha\, \delta(x - x_0)\, \delta_{\nu\nu_0},
\end{equation}
where $x_0$ and $y_0 = W\nu_0/N$ specify the impurity position. Generically,
$\alpha$ is a $4 \times 4$ matrix in sublattice and valley spaces. Below we
concentrate on the case of scalar potential with $\alpha$ being a number.

In order to study transport properties, we perform a standard unitary rotation
\cite{Titov07} of the Hamiltonian, $H \mapsto \mathcal{L}^\dagger H \mathcal{L}$
with $\mathcal{L} = (\sigma_x + \sigma_z)/\sqrt{2}$. After this transformation
the upper (lower) element of the spinor wave function represents right- (left-)
propagating mode in the leads. We then perform the discrete Fourier transform
with respect to $\nu$ and arrive at the Dirac equation
\begin{equation*}
 \frac{\partial \Psi(x)}{\partial x}
  = \left[
      \sigma_x \hat q - i \sigma_z \hat U(x)
    \right] \Psi(x),
\end{equation*}
where $\hat q$ is a diagonal matrix with entries $q_n = 2 \pi n / W$ being the
discrete transverse momenta and $n = -N/2, \ldots, N/2$ (for definiteness, we
assume periodic boundary conditions). The impurity potential is represented by
the operator $\hat U(x) = \alpha \delta(x - x_0) | \Phi \rangle \langle \Phi |$
that projects onto the state with the wave function $\Phi(q_n) = N^{-1/2} e^{i
q_n y_0}$. Using this separable form of the impurity potential $\hat U$, we can
explicitly calculate the transfer matrix $\mathcal{T}$ that relates the
wave-function amplitudes in the opposite leads, $\Psi(L) = \mathcal{T} \Psi(0)$.
The result is given by
\begin{equation*}
 \mathcal{T}
  = e^{\sigma_x \hat q (L - x_0)}
    e^{i \sigma_z\, \alpha | \Phi \rangle \langle \Phi |}
    e^{\sigma_x \hat q x_0}.
\end{equation*}
Inverting the element $\mathcal{T}_{11}$ (where indices refer to the $\sigma$
space), we obtain an exact expression for the transmission amplitude from the
$n$th channel in the left lead to the $m$th channel in the right lead (see
Appendix \ref{App:tm}),
\begin{equation}
 \label{tgeneral}
 t_{nm}
  = \frac{\delta_{nm}}{\cosh q_n L}
    -\frac{2\gamma_{nm} [ z + iN \cot(\alpha/2)]^{-1}}
      {\cosh{q_nL}\, \cosh{q_mL}},
\end{equation}
with $\gamma_{nm} = e^{i(q_n - q_m) y_0} \cosh[q_n x_0 - q_m(L - x_0)]$ and
\begin{equation}
 \label{z}
 z
  = \sum_n \cosh[q_n(L - 2 x_0)]/\cosh (q_n L).
\end{equation}

The first term in Eq.\ (\ref{tgeneral}) describes the clean system ($\alpha =
0$). The second term represents the effect of the impurity which is particularly
important in the vicinity of resonant values $\alpha_n = \pi (2n + 1)$. The
quantity $z$ defined by Eq.\ (\ref{z}) is $\pi W$ times the local density of
states of a clean system at the position of impurity.

\begin{figure}
 \centerline{\includegraphics[width=0.9\columnwidth]{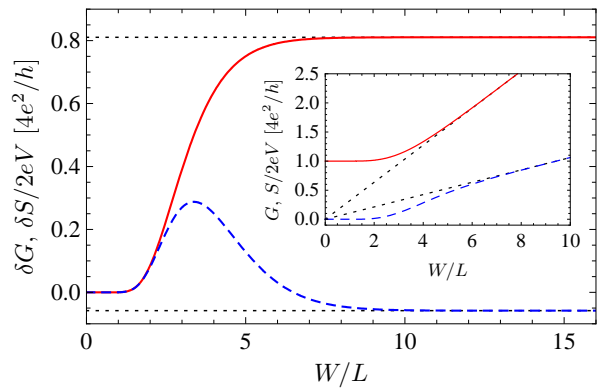}}
 \caption{(Color online) Contribution to the conductance $G$ (solid) and to the
shot noise $S$ (dashed) from a resonant impurity placed at $x_0 = L/2$, as a
function of the aspect ratio $W/L$. Dotted lines show asymptotic values from
Eq.\ (\protect\ref{resonance}). Inset: conductance and noise of a clean sample,
cf.\ Ref.\ \protect\cite{Tworzydlo06Beenakker08rev}.}
 \label{Fig:WL}
\end{figure}

The statistics of the charge transport is described by the cumulant generating
function $\mathcal{F}(\chi)$ \cite{LevitovLesovik} as $c_n = \lim_{\chi \to
0} \partial^n \mathcal{F} / \partial \chi^n$, where $c_1 = G/g_0$ is the
dimensionless conductance, $c_2 = S/2eV g_0$ is the dimensionless noise, etc.
For our purposes it is more convenient to use the variable $\phi$ defined via
$\chi = 2\ln|\cos(\phi/2)|$, which yields
\begin{equation}
 \label{F}
 \mathcal{F}(\phi)
  = \ln \det \left[
      1 - \hat{t} \hat{t}^\dagger\,\sin^2(\phi/2)
    \right].
\end{equation}
Using Eq.\ (\ref{tgeneral}), we calculate conductance and noise of the
sample with a resonant impurity placed in the center, see Fig.\ \ref{Fig:WL}.
The general expression for the generating function applicable for any
strength and position of the impurity is given in Appendix \ref{App:tm}.

In the limit $W\gg L$ the summations over transversal modes can be replaced by
momentum integrals. For the clean system this yields $\mathcal{F}_0 = - W
\phi^2/4\pi L$. The impurity correction to $\mathcal{F}$ takes the form
\begin{equation}
 \label{final}
 \delta\mathcal{F}
  = 2\ln \left[
      1 - \frac{\ell^2}{(4L)^2} \left(
        \frac{\phi^2}{\pi^2} - \frac{1}{\sin^2(\pi x_0/L)}
      \right)
    \right],
\end{equation}
where $\ell = 2\pi a \tan(\alpha/2)$ (as we show below, $\ell$ is the scattering
length at low energies). In particular, the correction to the conductance $G_0 =
g_0 W/\pi L$ reads
\begin{equation}
 \label{G}
 \delta G
  = (8 g_0/\pi^2) \left[
      \sin^{-2}(\pi x_0/L) + (4L)^2/\ell^2
    \right]^{-1}.
\end{equation}
This completes our analysis of the discretized model (\ref{discrete}).

Let us now turn to the microscopic description of the impurity within the
continuous Dirac model, assuming that the impurity potential is rotational
symmetric, $U(\mathbf{r}) = u(|\mathbf{r} - \mathbf{r}_0|)$, where
$\mathbf{r}_0$ is the impurity position. To calculate FCS we use the matrix
Green function approach \cite{Nazarov94, Ludwig}. The Green function in the
retarded-advanced (RA) space satisfies the equation
\begin{equation*}
 \begin{pmatrix}
   \mu(x) - H + i0 & -\sigma_x \zeta \delta(x) \\
   -\sigma_x \zeta \delta(x-L) & \mu(x) - H - i0
 \end{pmatrix} \check{G}(\mathbf{r}, \mathbf{r}')
  = \delta(\mathbf{r}-\mathbf{r}'),
\end{equation*}
where $\zeta = \sin(\phi/2)$ is the counting field and $\mu$ is the chemical
potential which is zero inside the sample and infinite in the leads. An
explicit solution to the above equation is given in Appendix \ref{App:green}.

The generating function can be expressed through $\check G$ as
$\mathcal{F}(\phi) = \mathop{\mathrm{Tr}} \ln \check G^{-1}$, where
$\mathop{\mathrm{Tr}}$ is the full operator trace. Expanding $\mathcal{F}(\phi)$
in $U(\mathbf{r})$, we obtain the impurity correction to the generating
function as a series
\begin{equation}
 \label{loop}
 \delta\mathcal{F}
  = -\sum_{p=1}^\infty \frac{1}{p} \mathop{\mathrm{Tr}} \big(
      U \check{G}_0
    \big)^p,
\end{equation}
where $\check{G}_0$ is the Green function of the clean system. Evaluation of
Eq.\ (\ref{loop}) essentially simplifies if the impurity size $a$ is small
compared to $L$. In this limit the arguments of the Green function are close to
the position of the impurity, $\mathbf{r}_0$. Therefore it is useful to split
$\check{G}_0$ into the singular and regular parts, $\check{G}_0 = g +
\check{G}_\mathrm{reg}$, where
\begin{equation*}
 g(\mathbf{r}, \mathbf{r}')
  = -(i/2\pi)
    \bm{\sigma} \cdot (\mathbf{r} - \mathbf{r}')/|\mathbf{r} - \mathbf{r}'|^2
\end{equation*}
is the zero-energy Green function of the free Dirac fermion. The regular part
of the Green function, $\check{G}_\mathrm{reg}$, can be taken with equal
arguments $\mathbf{r} = \mathbf{r}' = \mathbf{r}_0$. Explicit expression for
$\check{G}_\mathrm{reg}(\mathbf{r})$ is given in Appendix \ref{App:green}. We
characterize the impurity by its $T$-matrix
\begin{equation*}
 T
  = \sum_{p = 1}^\infty \int d^2\mathbf{r}_1 \dots  d^2\mathbf{r}_p
    U(\mathbf{r}_1) g(\mathbf{r}_1, \mathbf{r}_2) \dots U(\mathbf{r}_p).
\end{equation*}
Rearranging singular and regular parts in Eq.\ (\ref{loop}) yields
\begin{equation*}
 \delta\mathcal{F}
  = \mathop{\mathrm{Tr}} \ln[1 - T \check G_\mathrm{reg}(\mathbf{r}_0)].
\end{equation*}
Note that this result holds for any $T$ matrix, including possible valley
mixing. In the case of rotational invariant impurity, the $T$ matrix reduces to
the scattering length $\ell$ in the \textit{s}-channel. Since
$\mathop{\mathrm{Tr}} \check G_\mathrm{reg}(\mathbf{r}_0) = 0$ and $
\check G_\mathrm{reg}^2(\mathbf{r}_0) = [\phi^2/\pi^2 - \sin^{-2} (\pi
x_0/L)]/(4L)^2$ (Appendix \ref{App:green}), the impurity correction to the
generating function reproduces the result (\ref{final}). This establishes a
relation between the phenomenological parameter $\alpha$ in Eq.\
(\ref{discrete}) and the actual profile of the impurity potential.

\begin{figure}
 \centerline{\includegraphics[width=0.8\columnwidth]{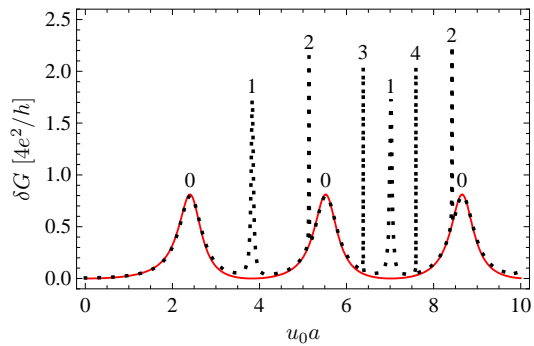}}
 \caption{(Color online) Correction to the conductance from a circular impurity
at $x_0 = L/2$ as a function of $u_0 a$. Solid line: analytic result, Eqs.\
(\protect\ref{G}), (\protect\ref{swave}); dotted line: numerical simulations
\protect\cite{Bardarson09} with the parameters $a/L = 0.2$, $W/L = 6$.
\vspace*{-0.35cm}}
 \label{Fig:Ua}
\end{figure}

It is convenient to derive the scattering length $\ell$ from the solution of the
Dirac equation in an infinite system \cite{HentschelNovikovBasko} at a low
energy $k$. Specifically, $\ell$ is related to the scattering cross-section
$\Lambda$ as $\Lambda = k \ell^2/2$. As an example, let us calculate $\ell$ for
an impurity with the rectangular profile $u(r) = u_0 \theta(a - r)$. Matching
the solutions of the free Dirac equation at $r < a$ and $r > a$, we obtain
\begin{equation}
 \label{swave}
 \ell
  = 2\pi a\, J_1(u_0 a) / J_0(u_0 a),
\end{equation}
where $J_n$ stands for the Bessel function. The scattering length diverges when
the impurity potential develops a quasibound state (one component of the spinor
wave function is localized) at zero energy. This implies resonant scattering,
which is a generic feature of any strong impurity independent of its profile.
In the quasiclassical limit, $u_0 a \gg 1$, one finds $\ell = 2\pi a \tan(u_0 a
- \pi/4)$, which corresponds to $\alpha = 2 u_0 a - \pi/2$ in Eq.\
(\ref{discrete}).

In Fig.\ \ref{Fig:Ua} we compare the results (\ref{G}), (\ref{swave}) to the
conductance calculated numerically in Ref.\ \cite{Bardarson09} for the case of
the disk impurity. In addition to the \textit{s}-wave resonances, that are
perfectly described by our theory, there exist higher resonances that can be
included into our consideration in a similar manner, see Appendix
\ref{App:pwave}. The position of the resonances are given by the zeroes of
$J_m(u_0 a)$, where the corresponding index $m = 0, 1, 2, \ldots$ is specified
in Fig.\ \ref{Fig:Ua}. The resonance widths scale as $(a/L)^{2m + 1}$.

\begin{figure}
 \centerline{\includegraphics[width=0.9\columnwidth]{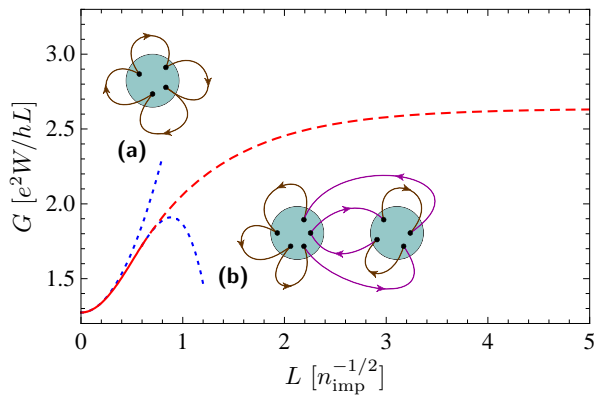}}
 \caption{(Color online) Conductance vs. $L$ for resonant impurities. Dotted
lines: (a) one- and (b) two-impurity contributions. Inset: corresponding typical
diagrams. Dashed line illustrates schematically the crossover to diffusive
regime.
\vspace*{-0.35cm}}
 \label{Fig:diagr}
\end{figure}

Now we turn to the case of disordered graphene with a small concentration
of impurities randomly distributed  over the sample. The average generating
function $\bar{\mathcal{F}}$ can be found to the linear order in
$n_\mathrm{imp}$ (see Fig.\ \ref{Fig:diagr}a) by the integration of Eq.\
(\ref{final}) over the sample area,
\begin{equation*}
 \bar{\mathcal{F}}
  = -\frac{W \phi^2}{4\pi L} + 4 N_\mathrm{imp} \ln \left(
      1 + \sqrt{1 + \frac{16L^2}{\ell^2} - \frac{\phi^2}{\pi^2}}
    \right),
\end{equation*}
where $N_\mathrm{imp} = n_\mathrm{imp} WL$ is the total number of impurities.
Resonant impurities ($\ell=\infty$) contribute on average the universal
correction (\ref{resonance}) to the conductance and noise, yielding the
conductivity $\sigma = GL/W$ and Fano factor
\begin{equation*}
 \sigma
  = \frac{4e^2}{\pi h} \left(
      1 + \frac{4}{\pi}\, n_\mathrm{imp} L^2
    \right),
 \qquad
 F
  = \frac{1}{3} - \frac{12}{\pi^3}\, n_\mathrm{imp} L^2.
\end{equation*}
Away from a resonance, $a \ll \ell \ll L$, we find
\begin{equation}
 \sigma
  = \frac{4e^2}{\pi h} \left(
      1 + \frac{n_\mathrm{imp} \ell^2}{2\pi}
    \right),
 \qquad
 F
  = \frac{1}{3} - \frac{n_\mathrm{imp} \ell^4}{8 \pi^3 L^2}.
 \label{weak}
\end{equation}

To establish the limits of validity of these ballistic results, we estimate the
contribution of the order $n_\mathrm{imp}^2$ in the virial expansion. The
diagrams involving two resonant impurities, Fig.\ \ref{Fig:diagr}b, yield the
free-energy correction
\begin{equation*}
 \delta^{(2)}\mathcal{F}
  = \frac{1}{2}\mathop{\mathrm{Tr}} \ln\big[
      1 - \check G_\mathrm{reg}^{-1}(\mathbf{r}_1)
      \check G_0(\mathbf{r}_1, \mathbf{r}_2)
      \check G_\mathrm{reg}^{-1}(\mathbf{r}_2)
      \check G_0(\mathbf{r}_2, \mathbf{r}_1)
    \big],
\end{equation*}
which is a contribution of atypical pairs of scatterers separated by a distance
$\lesssim L$. The corresponding corrections to the conductance and Fano factor
are (Appendix \ref{App:twoimps})
\begin{equation*}
 \delta^{(2)}\sigma
  = -1.032\, \frac{e^2}{h} n^2_\mathrm{imp} L^4,
 \qquad
 \delta^{(2)} F
  = 0.8623\, n^2_\mathrm{imp} L^4.
\end{equation*}
Two closely located strong impurities detune each other from the resonance,
hence the negative correction to the conductance. For off-resonance impurities,
$\ell \ll L$, the multiple scattering correction to conductance can be derived
from the first-order result Eq.\ (\ref{weak}) supplied by the logarithmic
renormalization of the effective impurity strength $\alpha_0 = n_\mathrm{imp}
\ell^2/2\pi$ according to Refs.\ \cite{OurPRB, Schuessler09},
\begin{equation*}
 \sigma
  = \frac{4e^2}{\pi h} \left(
      1 + \frac{n_\mathrm{imp} \ell^2}{2[ \pi - n_\mathrm{imp} \ell^2
\log(L/\ell)]}
    \right).
\end{equation*}

When $\delta\sigma \sim e^2/h$ (which corresponds to mean free path $\sim L$),
the system enters the diffusive regime, Fig.\ \ref{Fig:diagr}. This happens at
$n_\mathrm{imp} \min
\{ L^2, \ell^2 |\log(L/\ell)| \} \sim 1$. Remarkably, for resonant impurities
this implies just one impurity per square $L \times L$. The scaling of
conductivity in the diffusive regime is determined by the symmetries of
impurities in sublattice and valley space \cite{OurPapers}. In particular, for
vacancies the quantum interference is suppressed in view of chiral symmetry
\cite{OurPRB}. This should yield the universal (as long as $N_\mathrm{imp}$ is
small compared to the number of carbon atoms) minimal conductivity $\sim e^2/h$.

The situation with strong potential scatterers (e.g., adsorbants inducing midgap
states) is similar for intermediate $L$ where the chiral symmetry is
approximately preserved, Fig.\ \ref{Fig:diagr}. At larger scales, $L > L_c$,
violation of this symmetry becomes important, and quantum interference comes
into play. The symmetry-breaking length $L_c$ is given by $L_c \sim
n_\mathrm{imp}^{-1/2} \min \{n_\mathrm{imp}\ell^2, 1/n_\mathrm{imp}a^2\}$: the
chiral symmetry is broken due to detuning from resonance as well as due to
higher (non-resonant) scattering channels. For impurities that do not mix the
valleys (i.e. smooth on the scale of the lattice constant) this implies
antilocalization (symplectic symmetry class) leading to logarithmic increase of
conductivity with $L$. The valley mixing restores conventional (orthogonal) 2D
localization.

In conclusion, we have computed FCS for charge transport through an undoped
graphene sheet in the presence of strong potential impurities. The impurity
correction to FCS, Eq.\ (\ref{final}), is determined by the position of impurity
and the low-energy scattering length. The latter diverges when the impurity
potential has a quasibound state at zero energy. At such resonant conditions the
impurity correction becomes universal, Eq.\ (\ref{resonance}). Our results are
fully supported by numerical simulations, Fig.\ \ref{Fig:Ua}.

We are grateful to J.\ Bardarson and T.\ O.\ Wehling for valuable discussions.
We thank Centro de Ciencias de Benasque, Spain where this work was completed.
The work was supported by the EUROHORCS/ESF EURYI Award (I.V.G.) and Rosnauka
grant 02.740.11.5072.

\onecolumngrid

\appendix

\makeatletter
\c@secnumdepth=1
\makeatother

\section{Transfer matrix approach}
\label{App:tm}

The full counting statistics of the charge transport is encoded in the cumulant
generating function given by the formula
\begin{equation}
 \mathcal{F}(\phi)
  = \ln \det \left[
      1 - \hat t \hat t^\dagger\, \sin^2(\phi/2)
    \right],
 \label{Levitov}
\end{equation}
where $\hat t$ is an $N \times N$ matrix of transmission amplitudes, which are
defined with respect to the scattering states in the leads. The matrix $\hat t$
is obtained by inverting the block $\mathcal{T}_{11}$ of the transfer matrix
$\mathcal{T}$.

For the case of a single scalar impurity, the $2N \times 2N$ transfer matrix is
given by the product of three matrix exponents \cite{Titov07}
\begin{equation}
 \mathcal{T}
  = e^{\sigma_x \hat q (L - x_0)}
    e^{i\alpha \sigma_z\,|\Phi\rangle \langle\Phi|}
    e^{\sigma_x \hat q x_0},
 \label{TransferM}
\end{equation}
where the coordinate $x_0$ specifies the position of the impurity in $x$
direction. The expression (\ref{TransferM}) relays upon the sharp boundary
conditions at the metal-graphene interfaces: $x = 0$ and $x = L$. For periodic
boundary conditions in $y$, the momentum quantization gives $q_n = \pi(2n - N -
1)/W$, where $n = 1, 2, \dots, N$ (for simplicity, the number of channels, $N$,
is taken to be an odd number), and the matrix $\hat q$ is defined as $\hat q =
\mathop{\mathrm{diag}}(q_1, q_2, q_3, \dots, q_N)$. It is already assumed in the
construction of Eq.\ (\ref{TransferM}) that the impurity is point-like in $x$
direction. Furthermore, if the impurity is point-like in $y$ direction the
vector $|\Phi\rangle$ has the elements $\Phi(q_n) = N^{-1/2} e^{i q_n y_0}$,
where $y_0$ specifies the impurity position in $y$ direction. The impurity
strength in this phenomenological model is characterized by the dimensionless
parameter $\alpha$.

In the limit of point-like impurity one can use the projective property of the
operator $|\Phi\rangle\langle\Phi|$ in order to write
\begin{equation}
 e^{i\alpha \sigma_z\, |\Phi\rangle \langle\Phi|}
  = 1 - \left(
      1 - e^{i\alpha \sigma_z}
    \right)\, |\Phi\rangle \langle\Phi|.
\end{equation}
Then the expression (\ref{TransferM}) for the transfer matrix can be recast in
the following form:
\begin{equation}
 \mathcal T
  = e^{\sigma_x \hat q L} - e^{\sigma_x \hat q (L - x_0)}
    D |\Phi\rangle \langle\Phi|
    e^{\sigma_x \hat q x_0},
\end{equation}
where we have introduced the abbreviation $D = 1 - \exp{(i\alpha \sigma_z)}$. It
is also convenient to take advantage of $2 \times N$ matrix
\begin{equation}
 W_x
  = \frac{1}{\sqrt{N}} \begin{pmatrix}
      \cosh q_1 x  & \cosh q_2 x & \cosh q_3 x & \dots & \cosh q_N x\\
      \sinh q_1 x  & \sinh q_2 x & \sinh q_3 x & \dots & \sinh q_N x
    \end{pmatrix} e^{i \hat q y_0},
\end{equation}
which helps rewriting the matrix of transmission amplitudes, $\hat t =
(\mathcal{T}_{11})^{-1}$, as
\begin{equation}
 \hat t
  = \frac{1}{\cosh \hat q L - W_{L-x_0}^\dagger D W_{x_0}}
  = \frac{1}{\cosh \hat q L}
    +\frac{1}{\cosh\hat{q}L} W_{L-x_0}^\dagger
      \frac{1}{D^{-1} - W_{x_0} \frac{1}{\cosh \hat q L} W_{L - x_0}^\dagger}
      W_{x_0} \frac{1}{\cosh \hat{q}L} .
 \label{treduction}
\end{equation}
Thus the inversion of $N\times N$ block of the transfer matrix is reduced to the
inversion of $2 \times 2$ matrix in the last expression. Indeed, the
straightforward calculation yields
\begin{equation}
 W_{x_0} \frac{1}{\cosh \hat q L} W_{L - x_0}^\dagger
  = \frac{1}{2} \left(
      1 + \frac{z}{N}\sigma_z
    \right),
 \qquad\qquad
 z
  = \sum\limits_n \frac{\cosh q_n (L - 2x_0)}{\cosh q_n L}.
\end{equation}
Substituting this result into Eq.\ (\ref{treduction}) we obtain
\begin{equation}
 \hat t
  = \frac{1}{\cosh\hat q L}
    -\frac{1}{\cosh\hat q L} W_{L - x_0}^\dagger K W_{x_0}
      \frac{1}{\cosh\hat q L},
 \qquad\qquad
 K
  = \frac{2}{1 + (z/N)\sigma_z - 2D^{-1}}
  = \frac{2N \sigma_z}{z - iN \cot(\alpha/2)}.
 \label{tresult}
\end{equation}
This expression is equivalent to Eq.\ (\ref{tgeneral}). Using Eq.\
(\ref{tresult}) one can write the matrix product $\hat t ^\dagger \hat t $ as
\begin{equation}
 \hat t^\dagger \hat t
  = \frac{1}{\cosh\hat q L} \left(
      1 + \tilde W^\dagger \Sigma \tilde W
    \right) \frac{1}{\cosh\hat q L},
 \label{tildeW}
\end{equation}
where $\tilde W$ is a rectangular matrix of the dimension $4 \times N$ and $4
\times 4$ matrix $\Sigma$ is responsible for the impurity correction to the
transmission probabilities,
\begin{equation}
 \tilde W
  = \begin{pmatrix}
      W_{x_0} \\[4pt]
      W_{L-x_0} \dfrac{1}{\cosh \hat q L}
    \end{pmatrix},
 \qquad\qquad
 \Sigma
  = \begin{pmatrix}
      K^\dagger W_{L - x_0} \dfrac{1}{\cosh^2 \hat q L} W^\dagger_{L - x_0}K
       & -K^\dagger \\[8pt]
      -K & 0
    \end{pmatrix}.
\end{equation}

Using the decomposition (\ref{tildeW}) we can essentially simplify the
calculation of the determinant in Eq.\ (\ref{Levitov}) for the full counting
statistics. First, we factorize the determinant into the product of two
determinants: one for the full counting statistics of a clean system and the
other one for the impurity correction,
\begin{equation}
 e^{\mathcal{F}}
  = \det \left(
      1 - \frac{\sin^2(\phi/2)}{\cosh^2 \hat q L}
    \right) \det \left(
      1 - \frac{\sin^2(\phi/2)}{\cosh^2 \hat q L - \sin^2(\phi/2)}
        \tilde W^\dagger \Sigma \tilde W
    \right).
 \label{expF}
\end{equation}
The first determinant of the diagonal matrix is evaluated straightforwardly. In
the second one we perform a cyclic permutation of matrices and reduce it from $N
\times N$ down to $4 \times 4$. This matrix is diagonal in $\sigma$ space and
hence the determinant further reduces to the product of $2 \times 2$
determinants. After some algebraic manipulations we find out that these two
latter determinants are identical. This allows us to represent the generating
function $\mathcal{F}$ as a sum $\mathcal{F} = \mathcal{F}_0 +
\delta\mathcal{F}$, where $\mathcal{F}_0$ describes the clean system and
$\delta\mathcal{F}$ provides the impurity contribution to the full counting
statistics,
\begin{equation}
 \mathcal{F}_0
  = \sum_n \ln \left[
      1 - \frac{\sin^2(\phi/2)}{\cosh^2 q_n L}
    \right],
 \qquad\qquad
 \delta\mathcal{F}
  = 2 \ln \left[
      1 + \frac{P(\phi) - R_0^2 \sin^2\phi}{z^2 + N^2 \cot^2(\alpha/2)}
    \right].
 \label{deltaF}
\end{equation}
Here we have introduced the following notations:
\begin{equation}
 R_x
  = \frac{1}{2} \sum_n \frac{\cosh(2 q_n x)}{\cosh^2(q_n L) - \sin^2(\phi/2)},
 \qquad\qquad
 P(\phi)
  = R_{L - x_0}^2 + 2R_{L - x_0} R_{x_0} \cos\phi + R_{x_0}^2 - z^2.
\end{equation}
The generating function $\mathcal{F} = \mathcal{F}_0 + \delta\mathcal{F}$ given
by Eq.\ (\ref{deltaF}) is a general solution of the single-impurity problem
applicable for any strength and position of the impurity and arbitrary aspect
ratio of the sample.

In the limit $W \gg L$ the summations over transversal modes can be replaced by
momentum integrals, so that
\begin{equation}
 R_x
  = \frac{W}{2L}\, \frac{\sin(\phi x/L)}{\sin(\pi x/L) \sin\phi},
 \qquad
 R_0
  = \frac{W \phi}{2\pi L\, \sin\phi},
 \qquad
 z
  = \frac{W}{2L\, \sin(\pi x_0/L)},
 \qquad
 P(\phi)
  = 0.
\end{equation}
As a result, one obtains Eq.\ (\ref{final}).

\section{Matrix Green function}
\label{App:green}

The full counting statistics of the electron transport is conveniently
expressed in terms of the matrix Green function \cite{Nazarov94} in the
external counting field $\zeta = \sin(\phi/2)$. For the clean graphene sample
this Green function satisfies the following equation in the retarded-advanced
space:
\begin{equation}
 \label{keldysh}
 \begin{pmatrix}
   \mu(x) - \bm{\sigma} \mathbf{p} + i0 & -\sigma_x \zeta \delta(x) \\
   -\sigma_x \zeta \delta(x-L) & \mu(x) - \bm{\sigma} \mathbf{p} - i0
 \end{pmatrix} \check{G}_0 (\mathbf{r}, \mathbf{r}')
  = \delta(\mathbf{r} - \mathbf{r}'),
 \qquad\qquad
 \mu(x)
  = \begin{cases}
      0, & 0 < x < L, \\
      +\infty, & \text{$x < 0$ or $x > L$}.
    \end{cases}
\end{equation}
Since the operator in the left-hand side of the above equation commutes with the
$y$ component of the momentum, we will first calculate the Green function in the
mixed coordinate-momentum representation, $\check G_p (x, x')$. Inside the
sample this function satisfies
\begin{equation}
 \label{eqGreen}
 \left[
   i \sigma_x \frac{\partial}{\partial x} - \sigma_y p
 \right] \check G_p (x, x')
  = \delta(x - x').
\end{equation}

We will look for a general solution of this equation in the form
\begin{equation}
 \check G_p (x, x')
  = e^{\sigma_z p (x - L/2)} M e^{\sigma_z p (x' - L/2)},
 \qquad\qquad
 M
  = \begin{cases}
      M_<, & x < x', \\
      M_>, & x > x'.   
    \end{cases}
\end{equation}
The chemical potential profile together with the infinitesimal terms $\pm i0$
in Eq.\ (\ref{keldysh}) defines the boundary conditions for the Green function.
The counting field $\zeta$ can also be incorporated into the boundary
conditions. In terms of $M_\lessgtr$ we thus obtain
\begin{equation}
 \label{boundary}
 \begin{pmatrix}
   1 & 1 & i\zeta & i\zeta \\
   0 & 0 & 1 & -1
 \end{pmatrix} e^{-\sigma_z p L/2} M_<
  = 0,
 \qquad
 \begin{pmatrix}
   1 & -1 & 0 & 0 \\
   -i\zeta & -i\zeta & 1 & 1
 \end{pmatrix} e^{\sigma_z p L/2} M_>
  = 0.
\end{equation}
Delta function in the right-hand side of Eq.\ (\ref{eqGreen}) yields a jump of
the Green function at $x = x'$ which provides the relation
\begin{equation}
 \label{jump}
 M_> - M_<
  = -i \sigma_x.
\end{equation}
The matrices $M_\lessgtr$, and hence the Green function, are completely
determined by Eqs.\ (\ref{boundary}, \ref{jump}),
\begin{equation}
 M_\lessgtr
  = \frac{-i}{2(\cosh^2 pL - \zeta^2)}\begin{pmatrix}
        \cosh pL & \zeta^2 - \dfrac{\sinh 2pL}{2} & i\zeta e^{-pL} & i\zeta\\
        \zeta^2 + \dfrac{\sinh 2pL}{2} & \cosh pL & i\zeta & i\zeta e^{pL} \\
        i\zeta e^{pL} & i\zeta & -\cosh pL & -\zeta^2 - \dfrac{\sinh 2pL}{2} \\
        i\zeta & i\zeta e^{-pL} & -\zeta^2 + \dfrac{\sinh 2pL}{2} & -\cosh pL
    \end{pmatrix} \pm \frac{i\sigma_x}{2}.
\end{equation}

Fourier transform in $p$ yields the Green function in the full coordinate
representation. To facilitate further calculations, we decompose this Green
function into the following product of matrices:
\begin{gather}
 \check G_0 (x, x'; y)
  = \frac{1}{4}\, \check V(x) \check \Lambda \begin{pmatrix}
      i \cosh\dfrac{\phi y}{2L} & \sinh\dfrac{\phi y}{2L} \\[8pt]
      \sinh\dfrac{\phi y}{2L} & -i \cosh\dfrac{\phi y}{2L}
    \end{pmatrix}_{RA} \begin{pmatrix}
      \dfrac{1}{\sin \frac{\pi}{2L} (x + x' + iy)} &
      \dfrac{1}{\sin \frac{\pi}{2L} (x - x' + iy)} \\[8pt]
      \dfrac{1}{\sin \frac{\pi}{2L} (x - x' - iy)} &
      \dfrac{1}{\sin \frac{\pi}{2L} (x + x' - iy)}
    \end{pmatrix}_\sigma \check \Lambda \check V^{-1}(x'), \label{G0} \\
 \check \Lambda
  = \begin{pmatrix}
      \sigma_z & 0 \\
      0 & 1
    \end{pmatrix},
 \qquad\qquad
 \check V(x)
  = \begin{pmatrix}
      \sin\dfrac{\phi (L - x)}{2L} & \cos\dfrac{\phi (L - x)}{2L} \\[8pt]
      i\cos\dfrac{\phi x}{2L} & i \sin\dfrac{\phi x}{2L}
    \end{pmatrix}_{RA}.
\end{gather}
Here we have introduced the source angle $\phi$ defined by $\zeta =
\sin(\phi/2)$. The matrices $\check V(x)$ and $\check V^{-1}(x')$ operate in the
retarded-advanced space only and hence commute with any disorder operators
placed between the Green functions. As a result, factors $\check V$ and $\check
V^{-1}$ drop from expressions for any closed diagrams. The matrices $\check
\Lambda$ in the above equation allow us to decompose the Green function into a
direct product of the two operators acting in the RA space and in the sublattice
space. The matrix $\check \Lambda$ commutes only with the potential disorder and
must be retained as a part of the Green function in the general case.

The regularized Green function arising in the calculation of diagrams with
point-like impurities takes especially simple form when $\check V(x)$ and
$\check V^{-1}(x)$ are singled out,
\begin{equation}
 \check G_\mathrm{reg} (x)
  = \lim_{\substack{x' \to x\\[2pt] y \to 0}} \left[
      \check G(x, x'; y)
      +\frac{i}{2\pi}\, \frac{\sigma_x(x - x') + \sigma_y y}
        {(x - x')^2 + y^2}
    \right]
  = \frac{i}{4}\, \check V(x) \begin{pmatrix}
      \dfrac{1}{\sin\frac{\pi x}{L}} & -\sigma_x \dfrac{\phi}{\pi} \\[8pt]
      \sigma_x \dfrac{\phi}{\pi} & -\dfrac{1}{\sin\frac{\pi x}{L}}
    \end{pmatrix} \check V^{-1}(x).
 \label{Greg}
\end{equation}
It is worth noting that the above result does not depend on the order of taking
the limits.

\section{Resonances in $p$-wave scattering}
\label{App:pwave}

In order to include higher scattering resonances in our consideration, we take
into account small deviations of the regularized Green function endpoints from
the center of impurity. Expansion up to the first order in these deviations
allows for the resonances in the $p$ scattering channel. Let us define the
following matrix containing the regularized Green function and its various
derivatives:
\begin{equation}
 \check{\mathbb{G}}_\mathrm{reg} (\mathbf{r}_0)
  = \lim_{\substack{
      \mathbf{r} \to \mathbf{r}_0\\
      \mathbf{r}' \to \mathbf{r}_0
    }} \begin{bmatrix}
      1 & \partial_{x'} & \partial_{y'} \\
      \partial_x & \partial_x \partial_{x'} & \partial_x \partial_{y'} \\
      \partial_y & \partial_y \partial_{x'} & \partial_y \partial_{y'}
    \end{bmatrix} \Big[
      \check G_0(\mathbf{r}, \mathbf{r}') - g (\mathbf{r} - \mathbf{r}')
    \Big].
\end{equation}
Here and below we use the matrix notation with brackets rather than parentheses
to distinguish the newly introduced matrix from the internal structure of the
Green function in the RA and sublattice spaces. The $[1,1]$ component of
$\check{\mathbb{G}}_\mathrm{reg}$ coincides with the regularized Green function
$\check G_\mathrm{reg}$ from Eq.\ (\ref{Greg}).

The impurity is described by its $T$ matrix which is an operator with the
following kernel:
\begin{multline}
 T(\mathbf{r}, \mathbf{r}')
  = U(\mathbf{r}) \delta^{(2)} (\mathbf{r} - \mathbf{r}')
    + U(\mathbf{r}) g(\mathbf{r} - \mathbf{r}') U(\mathbf{r}') \\
    + \sum_{p=1}^\infty \int d^2 \mathbf{r}_1 \ldots d^2 \mathbf{r}_p\;
      U(\mathbf{r}) g(\mathbf{r} - \mathbf{r}_1) U(\mathbf{r}_1)
      g(\mathbf{r}_1 - \mathbf{r}_2) \ldots U(\mathbf{r}_p)
      g(\mathbf{r}_p - \mathbf{r}') U(\mathbf{r}').
\end{multline}
In the main text we have focused on the $s$-wave resonances and characterized
the impurity by its integrated $T$ matrix. In the simplest case of circular
potential impurity the integrated $T$ matrix is nothing but the scattering
length $\ell$ in the $s$ channel. Here we retain more information about the
shape of the impurity and introduce the following integral of the $T$ matrix:
\begin{equation}
 \mathbb{T}
  = \int d^2\mathbf{r}\, d^2\mathbf{r}' \begin{bmatrix}
      1 & x' - x_0 & y' - y_0 \\
      x - x_0 & (x - x_0)(x' - x_0) & (x - x_0) (y' - y_0) \\
      y - y_0 & (y - y_0)(x' - x_0) & (y - y_0) (y' - y_0)
    \end{bmatrix} T(\mathbf{r}, \mathbf{r}').
\end{equation}
Here and below we assume $\mathbf{r} = \{x, y\}$ and $\mathbf{r}' = \{x', y'\}$.
The $[1,1]$ component of $\mathbb{T}$ is the integrated $T$ matrix used in the
main text.

With the above-defined matrices $\check{\mathbb{G}}_\mathrm{reg}$ and
$\mathbb{T}$, we can express the single-impurity correction to the free energy
in the form
\begin{equation}
 \delta\mathcal{F}
  = \ln \det \big[
      1 - \mathbb{T} \check{\mathbb{G}}_\mathrm{reg}(\mathbf{r}_0)
    \big].
 \label{dFp}
\end{equation}
This expression includes the product of $[1,1]$ components of
$\check{\mathbb{G}}_\mathrm{reg}$ and $\mathbb{T}$, reproducing the result
obtained in the main text, along with the next terms of the Taylor expansion
of $\check G(\mathbf{r}, \mathbf{r}')$ in the vicinity of $\mathbf{r}_0$.

Let us now calculate the matrix $\check{\mathbb{G}}_\mathrm{reg}$. Taking
derivatives of $\check G(\mathbf{r}, \mathbf{r}')$ we reveal the following
internal structure of $\check{\mathbb{G}}_\mathrm{reg}$:
\begin{equation}
 \check{\mathbb{G}}_\mathrm{reg}(x)
  = \begin{bmatrix}
      1 & 0 \\
      0 & \sigma_y \\
      0 & \sigma_x
    \end{bmatrix} \begin{bmatrix}
      \check G_\mathrm{reg}(x) & \check G^v_\mathrm{reg}(x) \\
      \check G^v_\mathrm{reg}(x) & \check G^t_\mathrm{reg}(x)
    \end{bmatrix} \begin{bmatrix}
      1 & 0 & 0 \\
      0 & \sigma_y & \sigma_x
    \end{bmatrix}.
 \label{Greg_reduced}
\end{equation}
Here the element $\check G_\mathrm{reg}$ is given by Eq.\ (\ref{Greg}) and the
other elements are
\begin{gather}
 \check G^v_\mathrm{reg}(x)
  = \frac{\pi}{8L}\, V(x) \begin{pmatrix}
      \sigma_z \dfrac{3 \phi^2 - \pi^2}{6\pi^2}
      -i\sigma_y \dfrac{\cos\frac{\pi x}{L}}{\sin^3\frac{\pi x}{L}}
       & -i\sigma_y \dfrac{\phi}{\pi \sin\frac{\pi x}{L}} \\[8pt]
      -i\sigma_y \dfrac{\phi}{\pi \sin\frac{\pi x}{L}}
       & \sigma_z \dfrac{3\phi^2 - \pi^2}{6\pi^2}
        + i\sigma_y \dfrac{\cos\frac{\pi x}{L}}{\sin^3\frac{\pi x}{L}}
    \end{pmatrix} V^{-1}(x), \\
 \check G^t_\mathrm{reg}(x)
  = \frac{i}{8L^2}\, V(x) \begin{pmatrix}
      \dfrac{\pi^2}{\sin^3\frac{\pi x}{L}}
      -\dfrac{\pi^2 + \phi^2}{2\sin\frac{\pi x}{L}}
       & \dfrac{\pi \phi \cos\frac{\pi x}{L}}{\sin\frac{\pi x}{L}}
        -\sigma_x \phi \dfrac{\pi^2 - \phi^2}{6\pi} \\[8pt]
      \dfrac{\pi \phi \cos\frac{\pi x}{L}}{\sin\frac{\pi x}{L}}
      +\sigma_x \phi \dfrac{\pi^2 - \phi^2}{6\pi}
       & -\dfrac{\pi^2}{\sin^3\frac{\pi x}{L}}
        +\dfrac{\pi^2 + \phi^2}{2\sin\frac{\pi x}{L}}
    \end{pmatrix} V^{-1}(x).
\end{gather}
Taking the advantage of the representation (\ref{Greg_reduced}) we can reduce
the size of the matrix in Eq.\ (\ref{dFp}) by introducing
\begin{equation}
 \mathbf{T}
  = \begin{bmatrix}
      1 & 0 & 0 \\
      0 & \sigma_y & \sigma_x
    \end{bmatrix} \mathbb{T} \begin{bmatrix}
      1 & 0 \\
      0 & \sigma_y \\
      0 & \sigma_x
    \end{bmatrix}
  = \int d^2\mathbf{r}\, d^2\mathbf{r}' \begin{bmatrix}
      1 \\ \sigma_y (x - x_0) + \sigma_x(y - y_0)
    \end{bmatrix} T(\mathbf{r}, \mathbf{r}') \begin{bmatrix}
      1 \\ \sigma_y (x' - x_0) + \sigma_x(y' - y_0)
    \end{bmatrix}^\dagger.
\end{equation}
Now the correction to the free energy takes the simplified form
\begin{equation}
 \delta\mathcal{F}
  = \ln \det \left\{
      1 - \mathbf{T} \begin{bmatrix}
        \check G_\mathrm{reg}(x_0) & \check G^v_\mathrm{reg}(x_0) \\
        \check G^v_\mathrm{reg}(x_0) & \check G^t_\mathrm{reg}(x_0)
      \end{bmatrix}
    \right\}.
 \label{dFp_reduced}
\end{equation}

In order to obtain the final result from Eq.\ (\ref{dFp_reduced}) we have to
establish a relation between the matrix $\mathbf{T}$ and the solution of the
scattering problem with a given impurity profile. Let us place the impurity
at the origin $\mathbf{r}_0 = 0$ and consider the scattering state with the
following asymptotic behavior far from the impurity:
\begin{equation}
 \psi(\mathbf{r})
  = \frac{e^{i \mathbf{k} \mathbf{r}}}{\sqrt{2}} \begin{pmatrix}
      1 \\ e^{i \varphi_\mathbf{k}}
    \end{pmatrix}
    -f_k(\varphi_\mathbf{r}, \varphi_\mathbf{k}) \sqrt{\frac{k}{4 \pi r}}\;
      e^{i k r + i \pi/4}\begin{pmatrix}
        1 \\ e^{i \varphi_\mathbf{r}}
      \end{pmatrix},
 \qquad\qquad
 k r \gg 1.
\end{equation}
This expression contains an incoming plane wave with the momentum $\mathbf{k}$
and an outgoing spherical wave with the direction-dependent scattering amplitude
$f_k$. We use the notation $\varphi_\mathbf{k}$ ($\varphi_\mathbf{r}$) for the
angle between the vector $\mathbf{k}$ ($\mathbf{r}$) and the direction of $x$
axis.

The scattering amplitude is directly related to the matrix element of the $T$
matrix between the incoming and outgoing waves
\begin{equation}
 f_k(\varphi, \varphi')
  = \frac{1}{2} \int d^2\mathbf{r}\, d^2\mathbf{r}' \begin{pmatrix}
      1 & e^{-i\varphi}
    \end{pmatrix} T(\mathbf{r}, \mathbf{r}') \begin{pmatrix}
      1 \\ e^{i\varphi'}
    \end{pmatrix} \exp\big[ ik(
      x' \cos\varphi' + y' \sin\varphi'
      -x \cos\varphi - y \sin\varphi
    )\big].
\end{equation}
The integrals entering the matrix $\mathbf{T}$ can be expressed through the $k
\to 0$ limit of the scattering amplitude and its derivatives in the following
way:
\begin{equation}
 \mathbf{T}
  = \lim_{k \to 0} \int_0^{2\pi} \frac{d\varphi\, d\varphi'}{2 \pi^2}
    \begin{pmatrix}
      1 & e^{-i\varphi'} \\
      e^{i\varphi} & e^{i(\varphi - \varphi')}
    \end{pmatrix} \begin{bmatrix}
      1 & -4i \sin(2\varphi') \frac{\partial}{\partial k} \\
      4i \sin(2\varphi) \frac{\partial}{\partial k}
       & 8 \sin(2\varphi) \sin(2\varphi') \frac{\partial^2}{\partial k^2}
    \end{bmatrix} f_k(\varphi, \varphi').
 \label{Tf_relation}
\end{equation}
Once the scattering amplitude for a given impurity is known, one can use Eqs.\
(\ref{Tf_relation}) and (\ref{dFp_reduced}) to find the correction due to such
an impurity to the full counting statistics, including $s$- and $p$-wave
scattering channels. Below we apply this general result to the circular
potential impurity.

In the case of axially symmetric impurity, the scattering amplitude depends on
the difference $\varphi - \varphi'$ and Eq.\ (\ref{Tf_relation}) simplifies to
\begin{equation}
 \mathbf{T}
  = \lim_{k \to 0} \int_0^{2\pi} \frac{d\varphi}{\pi} \begin{pmatrix}
      1 & 0 \\
      0 & e^{i \varphi}
    \end{pmatrix} \begin{bmatrix}
      1 & 0 \\
      0 & 4 \cos(2\varphi) \frac{\partial^2}{\partial k^2}
    \end{bmatrix} f_k(\varphi).
 \label{bfT}
\end{equation}
We can get rid of the $\varphi$ integral and express the result in terms of the
scattering phases $\delta_m$ where $m$ is a half-integer number --- $z$
component of the total momentum (angular + spin) of the Dirac electron. The
scattering amplitude is the following Fourier series with the coefficients
determined by the scattering phases:
\begin{equation}
 f_k(\varphi)
  = \frac{i}{k} \sum_m \left(
      e^{2 i \delta_m} - 1
    \right) e^{i(m-1/2)\varphi}.
 \label{fk}
\end{equation}
In the case of potential impurity preserving the time-reversal invariance, the
scattering phases obey the symmetry relation $\delta_m = \delta_{-m}$. At low
energies they decay according to $\delta_m = O(k^{2|m|})$. Using these
properties, we substitute Eq.\ (\ref{fk}) into Eq.\ (\ref{bfT}) and obtain
\begin{equation}
 \mathbf{T}
  = \begin{bmatrix}
      \ell & 0 \\
      0 & \ell_1
    \end{bmatrix},
 \qquad\qquad
 \ell
  = -4 \lim_{k \to 0} \frac{\delta_{1/2}}{k},
 \qquad\qquad
 \ell_1
  = -16 \lim_{k \to 0} \frac{\delta_{3/2}}{k^3}.
\end{equation}
Note that, contrary to $\ell$, the parameter $\ell_1$ has a dimension of the
third power of length.

If the potential impurity has the rectangular profile with the height $u_0$ and
radius $a$, the Dirac equation is easily solved yielding the following
scattering parameters:
\begin{equation}
 \ell
  = 2 \pi a \frac{J_1(u_0 a)}{J_0(u_0 a)},
 \qquad\qquad
 \ell_1
  = 2 \pi a^3 \frac{J_2(u_0 a)}{J_1(u_0 a)}.
 \label{bessels}
\end{equation}

\begin{figure}
 \centerline{\includegraphics[width=0.6\columnwidth]{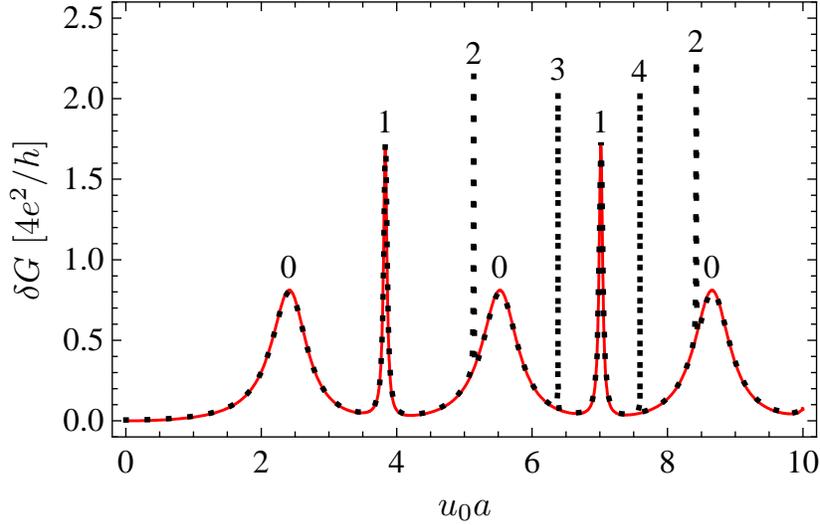}}
 \caption{Correction to the conductance from a circular impurity placed
at $x_0 = L/2$ as a function of $u_0 a$. Dotted line: numerical simulations
\protect\cite{Bardarson09} with the parameters $a/L = 0.2$, $W/L = 6$; solid
line: analytic result including $s$ and $p$ resonances.}
 \label{Fig:Ua2}
\end{figure}

Now we evaluate the determinant in Eq.\ (\ref{dFp_reduced}) and obtain the
result
\begin{multline}
 \delta \mathcal{F}
  = 2 \ln \Bigg\{
      1 - \frac{\ell^2}{16 L^2} \left(
        \frac{\phi^2}{\pi^2} - \frac{1}{\sin^2 \frac{\pi x_0}{L}}
      \right)
      -\frac{\pi^4 \ell_1^2}{256 L^6} \left[
        \left(
          \frac{\phi^2}{\pi^2} - 1
        \right)^2 \left(
          \frac{\phi^2}{9\pi^2} - \frac{1}{\sin^2 \frac{\pi x_0}{L}}
        \right) - \frac{4\cos^2 \frac{\pi x_0}{L}}{\sin^6 \frac{\pi x_0}{L}}
      \right] \\
      -\frac{\pi^2 \ell \ell_1}{128 L^4} \left[
        \left(
          \frac{\phi^2}{\pi^2} - \frac{1}{3}
        \right)^2 - \frac{4}{\sin^2 \frac{\pi x_0}{L}} \left(
          \frac{\phi^2}{\pi^2} + \cot^2 \frac{\pi x_0}{L}
        \right)
      \right] + O \big( \ell^2 \ell_1^2 \big)
    \Bigg\}.
 \label{sp-result}
\end{multline}
Here we have neglected the term $\sim \ell^2 \ell_1^2$ since it is always
much smaller than the term $\sim \ell \ell_1$. In Fig.\ \ref{Fig:Ua2} we compare
the conductance calculated from Eq.\ (\ref{sp-result}) as a function of $u_0 a$
with the results of numerical simulations \cite{Bardarson09}.

Exactly at the $p$ resonance, when $\ell_1 \to \infty$, correction to the
generating function becomes
\begin{equation}
 \delta \mathcal{F}
  = 2 \ln \left[
        \left(
          \frac{\phi^2}{\pi^2} - 1
        \right)^2 \left(
          \frac{\phi^2}{9\pi^2} - \frac{1}{\sin^2 \frac{\pi x}{L}}
        \right) - \frac{4\cos^2 \frac{\pi x}{L}}{\sin^6 \frac{\pi x}{L}}
      \right].
\end{equation}
From this expression we find the correction to the conductance and average it
with respect to the impurity position
\begin{equation}
 \delta G
  = -\frac{8 e^2}{h}\, \int \frac{dx}{L} \left.
      \frac{\partial^2 \delta \mathcal{F}}{\partial \phi^2}
    \right|_{\phi = 0}
  = \frac{32 e^2}{9\pi^2 h} \int \frac{dx}{L}
    \frac{\sin^4 \frac{\pi x}{L} \big (18 + \sin^2 \frac{\pi x}{L} \big)}
      {\big (2 - \sin^2 \frac{\pi x}{L} \big)^2}
  = \frac{16}{3\pi^2} \big( 15-8 \sqrt{2} \big) \frac{e^2}{h}
  \approx 1.992\; \frac{e^2}{h}.
\end{equation}
This value is bigger than the conductance correction at the $s$-wave resonance,
Eq.\ (\ref{resonance}).

\section{Two-impurity correction}
\label{App:twoimps}

\begin{figure}
 \centerline{\includegraphics[width=0.6\textwidth]{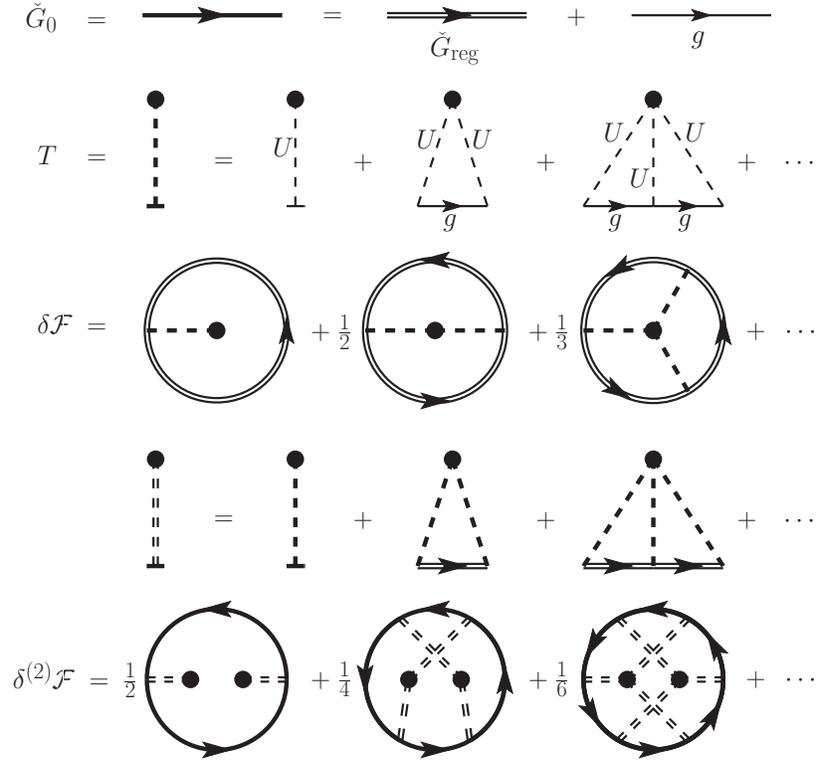}}
 \caption{Feynman diagrams for one- and two-impurity corrections to the free
energy.}
 \label{Fig:feynm}
\end{figure}

The second-order correction to the free energy involves the diagrams with two
impurities shown in Fig.\ \ref{Fig:feynm}. We sum up these diagrams assuming the
two impurities are located at the points $\{ x, y_0 \}$ and $\{ x', y_0 + y \}$
and obtain
\begin{equation}
 \delta^{(2)}\mathcal{F}
  = \frac{1}{2} \ln \det \Big\{
      1 - \big[1 - T \check G_\mathrm{reg}(x) \big]^{-1}
      T \check G(x, x'; y)
      \big[1 - T \check G_\mathrm{reg}(x') \big]^{-1}
      T \check G(x', x; -y)
    \Big\}.
\end{equation}
In the resonant case $T \to \infty$, this correction to the free energy takes
the form
\begin{equation}
 \delta^{(2)}\mathcal{F}
  = \frac{1}{2} \ln \det \Big[
      1 - \check G_\mathrm{reg}^{-1}(x)
      \check G(x, x'; y)
      \check G_\mathrm{reg}^{-1}(x')
      \check G(x', x; -y)
    \Big].
\end{equation}
With the Green functions calculated above, we evaluate the determinant and
obtain the result
\begin{gather}
 \delta^{(2)}\mathcal{F}
  = \ln\frac{
      \phi^4 - \phi^2 \big( a_+^2 + a_-^2 + b_+^2 + b_-^2 + c_+^2 + c_-^2 \big)
      + \big( a_+ a_- + b_+ b_- + c_+ c_- \big)^2
    }{(\phi^2 - a_+^2)(\phi^2 - a_-^2)}, \\
 a_\pm
  = \frac{\pi}{\sin \pi (u \pm v)},
 \qquad
 b_\pm
  = \frac{\pi}{\sin \pi (v \pm w)},
 \qquad
 c_\pm
  = \frac{\pi}{\sin \pi (w \pm u)}, \\
 u
  = \frac{x + x'}{2L},
 \qquad
 v
  = \frac{x - x'}{2L},
 \qquad
 w
  = \frac{iy}{2L}.
\end{gather}
This expression is to be averaged with respect to the impurity positions. To
calculate corrections to the conductance and noise, we expand the free energy in
small $\phi$ up to the forth order and integrate each term over $x$, $x'$, and
$y$ numerically. Together with the zeroth- and the first-order terms in
$n_\mathrm{imp}$, this yields
\begin{multline}
 \bar{\mathcal{F}}
  = -\frac{W \phi^2}{4\pi L}
    +4 n_\mathrm{imp} WL \ln \big( \pi+ \sqrt{\pi^2 - \phi^2} \big)
    +n_\mathrm{imp}^2 W \int_0^L dx\, dx' \int_{-\infty}^\infty dy\;
      \delta^{(2)}\mathcal{F} \\
  = \mathop{\mathrm{const}} - \frac{W}{L} \left[
      \frac{\phi^2}{4 \pi} + n_\mathrm{imp} L^2 \left(
        \frac{\phi^2}{\pi^2} + \frac{3 \phi^4}{8 \pi^4} + \ldots
      \right) - n_\mathrm{imp}^2 L^4 \left(
        0.1290\; \frac{\phi^2}{2} + 0.08823\; \frac{\phi^4}{24} + \ldots
      \right)
    \right].
\end{multline}
The conductance and the Fano factor up to the second-order corrections are
\begin{gather}
 G
  = -\frac{8 e^2}{h}\, \left.
      \frac{\partial^2 \bar{\mathcal{F}}}{\partial \phi^2}
    \right|_{\phi = 0}
  = \frac{e^2 W}{h L} \left[
      \frac{4}{\pi}
      +\frac{16}{\pi^2}\; n_\mathrm{imp} L^2
      -1.032\; n_\mathrm{imp}^2 L^4
    \right], \\
 F
  = \frac{1}{3} - \frac{2}{3}\,
    \frac{\partial^4 \bar{\mathcal{F}}/\partial \phi^4 |_{\phi = 0}}
    {\partial^2 \bar{\mathcal{F}}/\partial \phi^2 |_{\phi = 0}}
  = \frac{1}{3}
    -\frac{12}{\pi^3}\; n_\mathrm{imp} L^2
    +0.8623\; n_\mathrm{imp}^2 L^4.
\end{gather}


\vspace{-0.5cm}

\begin{thebibliography}{99}

\bibitem{Novoselov04}
\vspace{-0.5cm}
K.\,S.\,Novoselov \textit{et al.}, Science \textbf{306}, 666 (2004).

\bibitem{Novoselov05}
K.\,S.\,Novoselov \textit{et al.}, Nature (London) \textbf{438}, 197 (2005);
Y.\,Zhang \textit{et al.}, Nature (London) \textbf{438}, 201 (2005); Y.-W.\,Tan
\textit{et al.}, Eur.\ Phys.\ J.\ Spec.\ Top.\ \textbf{148}, 15 (2007).

\bibitem{GuineaRMP}
A.\,H.\,Castro Neto \textit{et al.}, Rev.\ Mod.\ Phys.\ \textbf{81}, 109 (2009).

\bibitem{Katsnelson}
M.\,I.\,Katsnelson, Eur.\ Phys.\ J.\ B \textbf{51}, 157 (2006).

\bibitem{Tworzydlo06Beenakker08rev}
J.\,Tworzyd{\l}o \textit{et al.}, Phys.\ Rev.\ Lett. \textbf{96}, 246802 (2006);
C.\,W.\,J.\,Beenakker, Rev.\ Mod.\ Phys. \textbf{80}, 1337 (2008).

\bibitem{Ludwig}
S.\,Ryu \textit{et al.}, Phys.\ Rev.\ B \textbf{75}, 205344 (2007).

\bibitem{Miao07}
F.\,Miao, \textit{et al.}, Science \textbf{317}, 1530 (2007).

\bibitem{Danneau08}
R.\,Danneau \textit{et al.}, Phys.\ Rev.\ Lett.\ \textbf{100}, 196802 (2008).

\bibitem{Titov07}
M.\,Titov, Europhys.\ Lett.\ \textbf{79}, 17004 (2007).

\bibitem{Schuessler09}
A.\,Schuessler \textit{et al.}, Phys.\ Rev.\ B \textbf{79}, 075405 (2009).

\bibitem{Bardarson07}
J.\,H.\,Bardarson, \textit{et al.}, Phys.\ Rev.\ Lett.\ \textbf{99}, 106801
(2007);
K.\,Nomura \textit{et al.}, \textit{ibid.} \textbf{99}, 146806 (2007);
P.\,San-Jose \textit{et al.}, Phys.\ Rev.\ B \textbf{76}, 195445 (2007);
C.\,H.\,Lewenkopf \textit{et al.}, \textit{ibid.} \textbf{77}, 081410R (2008);
J.\,Tworzydlo \textit{et al.}, \textit{ibid.} \textbf{78}, 235438 (2008).

\bibitem{OurPapers}
P.\,M.\,Ostrovsky \textit{et al.}, Phys.\ Rev.\ Lett.\ \textbf{98}, 256801
(2007); Eur.\ Phys.\ J.\ Spec.\ Top.\ \textbf{148}, 63 (2007).

\bibitem{OurPRB}
P.\,M.\,Ostrovsky \textit{et al.}, Phys.\ Rev.\ B \textbf{74}, 235443 (2006).

\bibitem{Guinea}
T.\,Stauber \textit{et al.}, Phys.\ Rev.\ B \textbf{76}, 205423 (2007).

\bibitem{Lichtenstein}
T.\ O.\ Wehling \textit{et al.}, Phys.\ Rev.\ B \textbf{75}, 125425 (2007).

\bibitem{Geim}
F.\ Schedin \textit{et al.}, Nat.\ Mater.\ \textbf{6}, 652 (2007).

\bibitem{graphane}
D.\,C.\,Elias \textit{et al.}, Science \textbf{323}, 610 (2009).

\bibitem{Bouchiat}
B.\ M.\ Kessler \textit{et al.}, arXiv:0907.3661.

\bibitem{Morpurgo}
H.\ B.\ Heersche \textit{et al.}, Nature \textbf{446}, 56 (2007).

\bibitem{Bolotin}
K.\ I.\ Bolotin \textit{et al.}, Solid State Commun. \textbf{146}, 351 (2008).

\bibitem{Andrei}
Xu Du \textit{et al.}, Int.\ J.\ Mod.\ Phys.\ B \textbf{22}, 4579 (2008).

\bibitem{Nazarov94}
Yu.\,V.\,Nazarov, Phys.\ Rev.\ Lett.\ \textbf{73}, 134 (1994).

\bibitem{LevitovLesovik}
L.S.Levitov and G.B.Lesovik, JETP Lett.\ \textbf{58}, 230 (1993).

\bibitem{HentschelNovikovBasko}
M.\,Hentschel and F.\,Guinea, Phys.\ Rev.\ B \textbf{76}, 115407 (2007);
D.\ S.\ Novikov, \textit{ibid.} \textbf{76}, 245435 (2007);
D.\,M.\,Basko, \textit{ibid.} \textbf{78}, 115432 (2008).

\bibitem{Bardarson09}
J.\,Bardarson \textit{et al.}, Phys.\ Rev.\ Lett.\ \textbf{102}, 226803 (2009).

\end{thebibliography}
\end{document}